\documentclass[%
 reprint,
superscriptaddress,
 amsmath,amssymb,
 aps,
prb,
floatfix,
]{revtex4-2}

\usepackage{ulem}
\usepackage{graphicx}
\usepackage{dcolumn}
\usepackage{bm}
\usepackage{xcolor}

\begin{document}

\preprint{}

\title{Pressure-induced insulator-to-metal transition in van der Waals compound CoPS$_3$}

\author{Takahiro Matsuoka}
\email[]{tmatsuok@utk.edu}
\affiliation{Department of Materials Science and Engineering, University of Tennessee, Knoxville, TN 37996, USA}

\author{Rahul Rao}
\affiliation{Materials and Manufacturing Directorate, Air Force Research Laboratory, Wright-Patterson Air Force Base, OH 45433, USA}

\author{Michael A. Susner}
\affiliation{Materials and Manufacturing Directorate, Air Force Research Laboratory, Wright-Patterson Air Force Base, OH 45433, USA}

\author{Benjamin S. Conner}
\affiliation{Sensors Directorate, Air Force Research Laboratory, Wright-Patterson Air Force Base, OH 45433, USA}
\affiliation{National Research Council, Washington D.C. 20001, USA}

\author{Dongzhou Zhang}
\affiliation{Hawaii Institute of Geophysics and Planetology, University of Hawaii at Manoa, 1680 East-West Road, Honolulu, HI 96822}
\affiliation{GSECARS, University of Chicago, 9700 S Cass Ave, Argonne, IL 60439}

\author{David Mandrus}
\email[]{dmandrus@utk.edu}
\affiliation{Department of Materials Science and Engineering, University of Tennessee, Knoxville, TN 37996, USA}
\affiliation{Department of Physics and Astronomy, University of Tennessee, Knoxville, TN 37996, USA}
\affiliation{Materials Science and Technology Division, Oak Ridge National Laboratory, Oak Ridge, TN 37831, USA}

\date{\today}

\begin{abstract}
We have studied the insulator-to-metal transition and crystal structure evolution under high pressure in the van der Waals compound CoPS$_3$ through \textit{in-situ} electrical resistance, Hall resistance, magnetoresistance, X-ray diffraction, and Raman scattering measurements. 
CoPS$_3$ exhibits a $C2/m$ $\rightarrow$ $P\overline{3}$ structural transformation at 7 GPa accompanied by a 2.9$\%$ reduction in the volume per formula unit.
Concomitantly, the electrical resistance decreases significantly, and CoPS$_3$ becomes metallic.
This metallic CoPS$_3$ is a hole-dominant conductor with multiple conduction bands. 
The linear magnetoresistance and the small volume collapse at the metallization suggest the incomplete high-spin $\rightarrow$ low-spin transition in the metallic phase. 
Thus, the metallic CoPS$_3$ possibly possesses an inhomogeneous magnetic moment distribution and short-range magnetic ordering.
This report summarizes the comprehensive phase diagram of $M$PS$_3$ ($M$ = V, Mn, Fe, Co, Ni, and Cd) that metalize under pressures.
\end{abstract}

\maketitle
\section{\label{sec:level1}INTRODUCTION}
The transition metal thiophosphates \textit{M}PS$_3$ (\textit{M} = V, Mn, Fe, Co, Ni, and Cd) compounds form a family of quasi-two-dimensional (2D) compounds. 
They are isostructural in a monoclinic $C2/m$ symmetry, with individual lamellae composed of slightly distorted octahedral sites circumscribed by the S atoms bordering the van der Waals gap which are, in turn, arranged in a honeycomb lattice. Of the octahedra, 2/3 can be described as a +2 metal cation in an $M$S$_6$ cage. The remaining 1/3 of the octahedra are filled with P-P dimers that form a [P$_2$S$_6$]$^4-$ anionic sublattice that charge balances the aforementioned metal cations. The anionic sublattice is common to all of these compounds; the cations in turn impart the various functionalities native to these systems, including magnetism \cite{Susner2017, Zhu2020}. 
In this particular subset of the metal thiophosphate family, all compounds form a Mott insulating state at low temperatures and exhibit 2D antiferromagnetic (AFM) behavior, except for the Cd, which in the +2 oxidation state has a closed $d$-shell \cite{Wildes2017, Coak2021, Wildes1994, Wildes2015, Kim2019b}. 
The 2D magnetism in these materials has attracted recent attention due to the ability to study the effects of extreme anisotropy in low dimensions. The \textit{M}PS$_3$ family thus offers an enticing materials platform to study novel magnetic phenomena in low-dimensional materials in addition to the promise for applications in magnetic and spintronic devices because they can be exfoliated down to thin films \cite{Li2013, Joy1992, Lancon2016, Wildes2006}.
Therefore, elucidating these interesting 2D magnetic materials' physical properties is vital for future applications. In particular, it is essential to consider the structural and magnetic changes induced by imparting pressure to these materials.

External pressure is an effective perturbation tool because van der Waals compounds are highly compressible, especially in the inter-layer direction. 
Thus far, the structural, magnetic, and electronic evolutions under compression have been extensively studied for the \textit{M}PS$_3$ (\textit{M}  = V, Mn, Fe, Ni, Cd) and their analogous selenophosphate counterparts \textit{M}PSe$_3$.
Researchers have commonly observed that these materials exhibit an insulator-to-metal transition at high pressures (12-28 GPa) \cite{Coak2019, Wang2016a, Wang2018a, Haines2018a, Matsuoka2021, Ma2021, Kim2019b, Harms2020}. 
Additionally, spin-crossover (high to low-spin state) occurs in FePS$_3$ (FePSe$_3$) and MnPS$_3$ (MnPSe$_3$) concomitantly with the insulator-to-metal transition \cite{Wang2016a, Wang2018a}. 
Finally, FePSe$_3$ becomes superconducting with a superconducting transition temperature of 2.5 K at 9 GPa (increasing to 5.5 K at 30 GPa) \cite{Wang2018a}. 
Evidence for a metallization has also been reported recently in a bimetallic metal thiophosphate, Cu-deficient CuInP$_2$S$_6$ \cite{rao2021pressure}.
External pressure induces inter-layer sliding transitions, followed by a 10-20$\%$ volume collapse across the insulator-to-metal transition at room temperature \cite{Coak2019, Harms2020, Wang2016a, Coak2021, Haines2018a, Niu2022, Harms2022a,Ma2021}.
Other transitions can be more subtle. 
For example, the \textit{M}PS$_3$ (\textit{M}  = Fe, Mn) and V$_{0.9}$PS$_3$ compounds change from $C2/m$ to a $C2/m$ with a different monoclinic angle ($\beta$) or a trigonal $P\overline{3}1m$ \cite{Coak2019, Ouvrard1985V, Ourvrard1985, Harms2020, Wang2016a,Coak2020, Wang2018a,Coak2021, Harms2022a}.
CdPS$_3$ changes from $C2/m$ to a trigonal $R\overline{3}$ \cite{Niu2022}.
Previous studies of NiPS$_3$ observed up to five high-pressure phases ($P\overline{3}$, $P\overline{3}m1$, $P3m1$, $P3$, and $P1$) between ambient pressure and 39 GPa, making this composition unique among these van der Waals gapped magnetic materials \cite{Harms2022}.
Clearly, pressure-driven structural phase transitions are critical as drivers of new states of matter with the potential to host emergent properties. 

Focusing on the last of these properties, the trigonal distortion present in the octahedra bounding the metal cations (where the trigonal axis is parallel to the stacking direction) affects the degeneracy of the energy states associated with the octahedral crystal field splitting, thus creating highly anisotropic effects in the magnetism of these compounds \cite{Susner2017} that in turn contribute to their interesting behavior. 
From a structural perspective, the magnetic \textit{M}PS$_3$ or \textit{M}PSe$_3$ can be grouped into three main categories. In all compounds, spins are pointed along the $c$-axis except for NiPS$_3$ and CoPS$_3$, where spins are pointed parallel and antiparallel to the $a$ direction. 

The first grouping is MnPS$_3$ and MnPSe$_3$; both of these compounds are co-linear antiferromagnets with propagation vectors of $\bm{q}$ $= [0\ 0\ 0]$ \cite{ressouche_magnetoelectric_2010, bhutani2020, wiedenmann_1981}. 
We note that other works state that $\bm{q} = [0\ 1\ 0]$ for MnPS$_3$ \cite{Wildes1994}.
The second main grouping contains CoPS$_3$ and NiPS$_3$; they exhibit $\bm{q} = [0\ 1\ 0]$ \cite{Wildes2015, Wildes2017}. In CoPS$_3$, though the orientation of the moments is mostly along the $a$-axis, a small component may be along the $c$-axis as well \cite{Wildes2017}. Finally, the third grouping comprises the compounds FePS$_3$ and FePSe$_3$. The sulfide has been shown to have an incommensurate $\bm{q} = [1/2\  1/2\  0.34]$ \cite{rule_single-crystal_2007} while the selenide has a vector of $\bm{q} = [1/2\  0\  1/2]$ \cite{bhutani2020, wiedenmann_1981}. The careful reader may note that V$_{0.8}$PS$_3$ \cite{Ichimura1991}, NiPSe$_3$ \cite{LeFlem1982}, and many of the quarternary magnetic compounds \cite{Susner2017} have yet to be fully characterized in terms of magnetic structure and may yield fruitful investigations themselves if suitably sized crystals can be synthesized.

To date, the effects of high pressure on CoPS$_3$ have not been experimentally reported, probably due to the significant difficulty in the synthesis and single crystal growth of this compound. 
CoPS$_3$ is antiferromagnetic at ambient pressure with a Neel temperature of \textit{T}$_N$ = 122 K and a Weiss temperature of $\theta$ $= -116$ K \cite{Ouvrard1982}. 
Its effective moment is 4.9 $\mu$$_B$, slightly larger than the expected value for a pure spin moment of a Co$^{2+}$ ($S = $$\frac{3}{2}$) cation \cite{Ouvrard1982}, implying some degree of orbital contribution to the magnetization. A first-principles calculation, the only dedicated article to the pressure effects in CoPS$_3$, predicts a pressure-driven isostructural Mott transition accompanied by a spin-crossover \cite{Gu2021}.

In this study, we successfully grew large single crystals of CoPS$_3$.
Our electrical resistance, Raman scattering, and X-ray diffraction (XRD) measurements reveal an insulator-to-metal transition around 7 GPa, coinciding with a $C2/m$ $\rightarrow$ $P\overline{3}$ structural transformation and a 2.9$\%$ reduction in the volume per formula unit. 
Hall effect measurements find the metallic phase is a hole-dominant conductor. 
The linear magnetic field dependence of the magnetoresistance, combined with the small volume collapse at the metallization, suggests an incomplete high-spin $\rightarrow$ low-spin crossover in the metallic phase. 
Thus, metallic CoPS$_3$ may possess an inhomogeneous magnetic moment distribution and short-range magnetic ordering due to the coexisting high- and low-spin Co$^{2+}$ ions.
This report summarizes the comprehensive phase diagram of $M$PS$_3$ that metalize under compression.

\section{\label{sec:level2}EXPERIMENTAL METHOD}

We synthesized single crystals of CoPS$_3$ using the general procedures outlined in Refs. \cite{Wildes2017, Susner2017}.
Co powder (Alfa Aesar Puratronic, ~22 mesh, 99.998\%, reduced), P chunks (Alfa Aesar Puratronic, 99.999\%), and S (Alfa Aesar Puratronic, 99.9995\%) were combined in a near-stoichiometric ratio to form CoPS$_3$ together with an appropriate quantity of I$_2$ as the vapor transport agent in a sealed quartz ampoule, heated to the reaction temperature, and held there for 4 days \cite{may_practical_2020}. 
Typical crystals were 4-6 mm in size along the \textit{a–b} planes with several exceeding 12 mm. 
Typical thicknesses were $<$ 0.5 mm.

We used diamond anvil cells (DACs) for the high-pressure application.
We loaded a small single crystal of CoPS$_3$ in a DAC and connected five electrical probes made of platinum (Pt) for the electrical resistance, magnetoresistance, and Hall resistance measurements.
The \textit{a–b} planes of the single crystals were laid on the diamond’s flat surface to achieve a quasi-uniaxial compression.
A pre-compressed sodium chloride (NaCl) flake was placed underneath the sample, serving as a pressure-transmitting medium and ensuring that the pressure distribution across the sample was as homogeneous as possible. 
Several tiny ruby chips (Cr: Al$_2$O$_3$) were added with the samples as a pressure standard \cite{Mao1986}.
For further details, see the Supplemental material \cite{supp}.

We performed Raman spectroscopy measurements on a CoPS$_3$ crystal compressed in a DAC, using a Renishaw inVia Raman microscope with a 632.8 nm excitation laser at room temperature.
The pressure-transmitting medium was a 4:1 ratio methanol/ethanol solution.
The power of the excitation laser was tuned to $\sim$1 $\mu$W to minimize heating.

We conducted the XRD measurements at beamline 13BM-C at the Advanced Photon Source (APS), Argonne National Laboratory, utilizing a focused (12 $\mu$m (H) $\times$ 18 $\mu$m (V)) X-ray beam (wavelength $=$ 0.4340 \AA) in all measurements. 
Potassium chloride (KCl) was the pressure-transmitting medium.
We collected diffraction data on a flat panel detector array (Dectris Pilatus 1M-F, pixel size: 172 $\times$ 172 $\mu$m$^2$) in the forward scattering geometry at room temperature. 
We used Dioptas for two-dimensional XRD data reduction \cite{Dioptas}. 
We obtained lattice constants ($a$, $b$, and $c$) by the least-square-fitting of peak positions using PDindexer \cite{Seto2010a}.

\section{\label{sec:level3}RESULTS}
\subsection{Resistance vs. Pressure and Temperature}
Fig. 1a shows the pressure dependence of the electrical resistance (\textit{R}$_{xx}$) of CoPS$_3$ at room temperature without applying an external magnetic field.
To estimate the figure of electrical resistivity ($\rho$$_{xx}$), we calculated $\rho$$_{xx}$ for sample \#1 using the area/length ratio, $wt/l$ where $w  = 55$ $\mu$m is the width of the sample, $t = 20$ $\mu$m is the thickness, and $l = 50$ $\mu$m is the separation between electrical leads measured before applying pressure.
Since the separation between electrical leads does not change significantly (Fig. 1a), we assume that most of the change in $\rho$$_{xx}$ comes from the reduction of $t$.
From the  XRD results (Fig. 2b) discussed later, the $c$-axis shrinks 20$\%$ from the ambient pressure to the highest pressure (17 GPa).
Although the error does not affect the conclusion of the current report, readers are noted that a maximum of 20$\%$ of error should be included in the $\rho$$_{xx}$ in Fig. 1a.
We also note that we obtained the pressure by averaging the pressure values measured at room temperature before and after each temperature cycle for the electrical resistance measurements.
The actual pressures at low temperatures are thought to be somewhat different from the indicated values.

The \textit{R}$_{xx}$ decreases into the measurable range of our transport measurement system (maximum 2 M$\Omega$) above 2 GPa.
The \textit{R}$_{xx}$ shows a significant decrease up to 7 GPa by three orders of magnitude. Then it abruptly decreases by five further orders of magnitude when applying an additional mere 1 GPa.
Above 10 GPa, \textit{R}$_{xx}$ becomes almost independent of the external pressure.
The $\rho$$_{xx}$ reaches around 200 $\mu$$\Omega$cm, suggesting a metallic conductivity.
The similarities in the trends from two independent samples, \#1 and \#2, show good reproducibility of the transport results (Fig. 1a).
The insets of Fig. 1a show microphotographs of CoPS$_3$ at 4.5 GPa and 12 GPa under epi-illumination.
Here we see that the light reflection increases with pressure. 
The sample at 12 GPa is shiny and comparable to the Pt metal of the electrical probes, thus indicating a major electronic transition.

Upon releasing the pressure, the $R$$_{xx}$ traces back the $R$$_{xx}$ vs. \textit{P} curve of the compression down to 8 GPa.
Below 8 GPa, the $R$$_{xx}$ stays much lower (one to 3 orders of magnitude) than the compression, showing a large hysteresis. 
However, at 0 GPa, the compressed and decompressed lines are extrapolated to merge.  
This large hysteresis between compression and decompression implies the presence of a first-order transition.

To see if there are any temperature-dependent resistance effects, in Fig. 1b (top panel), we show $R$$_{xx}$ vs. \textit{T} at various pressures (Sample \#1) without an external magnetic field.
The CoPS$_3$ exhibits insulating or semiconducting behavior at 7.4 GPa and also displays a negative slope (d$R$$_{xx}$/d\textit{T}).
The origin of the hump at 7.4 GPa and 50 K is as of yet unknown. 
Above 10 GPa, d$R$$_{xx}$/d\textit{T} becomes positive along the entirety of the probed temperature range, thus revealing that CoPS$_3$ is metallic under these pressures.
We see no superconducting transition down to 2 K.
When we plot the $R$ as a function of \textit{T}$^2$ (inset in the bottom panel of Fig. 1b), we see that d$R$$_{xx}$/d\textit{T} fits the Fermi-liquid theory at temperatures below 15 K. 
Thus, we conclude that CoPS$_{3}$ exhibits a pressure-induced insulator-to-metal transition around 7 GPa.
At 3.5 GPa on decompression, CoPS$_{3}$ exhibits an insulator behavior (Fig. 1b).

\begin{figure}
\includegraphics[scale=0.36]{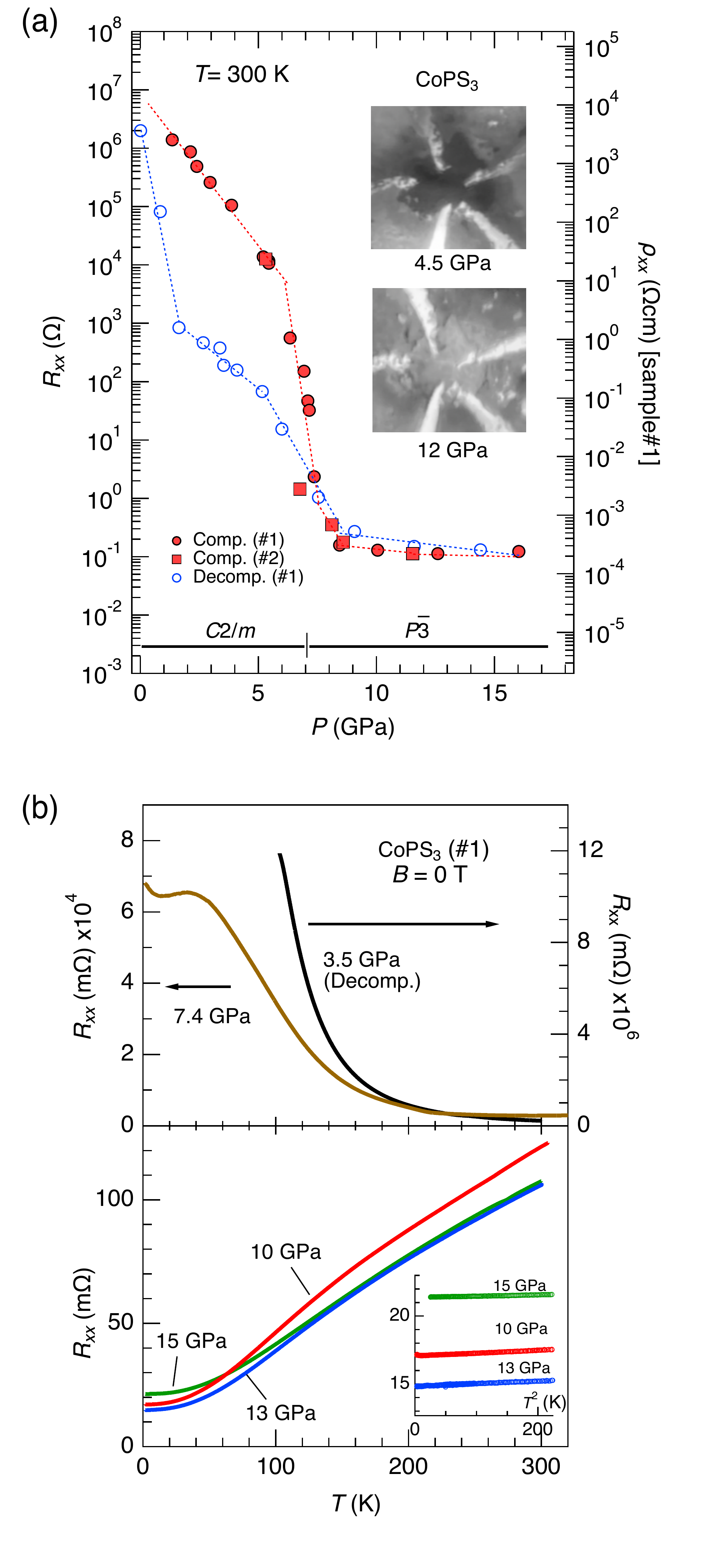}%
\caption{\label{fig1}(color online) (a) \textit{R}$_{xx}$ and $\rho$$_{xx}$ vs. \textit{P} at room temperature. The data of samples \#1 and \#2 are plotted for compression and decompression. The dotted lines are guides for the eyes. The vertical and horizontal solid lines indicate the structural phase diagram confirmed by our XRD and Raman scattering measurements. The inset pictures show sample $\sharp$1 at pressures. (b) $R$$_{xx}$ vs. \textit{T} at pressures obtained for Sample $\sharp$1 without an external magnetic field. All data plots are taken on warming. The inset figure plots the $R$$_{xx}$ vs. \textit{T}$^2$.}
\end{figure}

\subsection{XRD}
Figure 2a displays the representative XRD patterns of CoPS$_3$ at 0.5 and 17 GPa. 
We note that the observed XRD patterns are affected by the orientation reflecting the initial geometry of the single crystal when loaded in the DAC.
In our measurements, the \textit{a-b} plane of the layered structure is perpendicular to the X-ray beam. 
Although we rotated the DAC as much as allowed by the beamline instruments and the DAC opening angle, the diffraction peaks from ($hkl$) with large \textit{l} ($>$2) are invisible, limiting the number of peaks.
Additionally, the single crystal partially broke into several pieces during compression.
Therefore, the relative intensity between the diffraction peaks is inaccurate because the obtained XRD results are not in the form of even-intensity powder rings.
See the Supplementary Material for the XRD image recorded on a detector \cite{supp}.
At 0.5 GPa, the XRD peak positions agree well with the previously reported monoclinic $C2/m$ ($a$ = 5.844(1) \AA, $b$ = 10.127(1) \AA, $c$ = 6.562(4) \AA, $\beta$ = 107.04(2)$^{\circ}$) \cite{Wildes2017}.
At 17 GPa, we can index the XRD pattern to a trigonal structure with lattice constants $a = b =$ 5.570(5) \AA, $c =$ 5.13(2) \AA.
We name the high-pressure trigonal phase HP-I in this report.

Figure 2b shows the pressure dependencies of volume per formula unit (\textit{V}$_{f.u.}$) and the lattice constants, $a$, $b$, and $c$.
The \textit{V}$_{f.u.}$ is obtained by dividing a unit cell volume by the number of CoPS$_3$ units.
For the comparison between $C2/m$ and HP-I, we reduce the trigonal unit cell to a monoclinic unit cell ($\beta$ = 90$^{\circ}$) using the relation \textit{b}$_{mono.}$ $= 2a$$_{tri.}$$\times$cos30$^{\circ}$.
In agreement with the electrical transport measurements, the structure changes from $C2/m$ to HP-I at 7 GPa.
Concomitantly, the \textit{V}$_{f.u.}$ abruptly decreases by 2.9\% (84.690 \AA$^3$ $\rightarrow$ 82.269 \AA$^3$) at 7 GPa.
All the lattice constants $a$, $b$, and $c$ show a sharp discontinuity, with $c$ exhibiting the largest reduction ($\Delta$$c$ = 1.4 \AA).

Here, we focus on the observed reduction in the \textit{V}$_{f.u.}$.
Isostructural materials MnPS$_3$ and FePS$_3$ commonly collapse a volume by 10-20\% simultaneously with a spin-crossover, and the insulator-to-metal transition \cite{Haines2018a, Wang2016a, Wang2018a}.
This study has not performed a direct measurement, such as X-ray absorption and M\"{o}ssbauer spectroscopy, to investigate the electronic configuration of Co.
However, from the observed volume collapse and comparing FePS$_3$ and MnPS$_3$, it is reasonable to conclude that CoPS$_3$ exhibits the spin crossover ($S =$ 3/2 $\rightarrow$ 1/2) accompanied by the metallization at 7 GPa.

On the other hand, the ovserved volume reduction (2.9\%) of CoPS$_3$ is much smaller than MnPS$_3$ (19.7\%) and FePS$_3$ (10.6\%) \cite{Haines2018a, Wang2016a, Wang2018a}.
The ionic radii of high- (HS, 0.89 \AA{}) and low-spin (LS, 0.79 \AA{})  Co$^{2+}$ ions \cite{Shannon1976} make the HS $\rightarrow$ LS radius reduction by 11.2\%, which is not much smaller than that of Mn$^{2+}$ (HS: 0.97 \AA, LS: 0.81 \AA, 16.5\%) and Fe$^{2+}$ (HS: 0.92 \AA, LS: 0.75 \AA, 18.5\%) \cite{Shannon1976}.
Thus, the slight volume reduction of CoSP$_3$ cannot be explained simply by the difference between HS and LS radii.

In Fig. 2b, it is also noticeable that the $V$$_{f.u.}$ and $c$ in the HP-I phase show a steeper compression between 7 GPa and 12 GPa followed by moderate compression above 12 GPa, indicating a sign of $V$($P$) stabilization.
These series of changes in compression behavior suggest an electronic transition takes place in the HP-I phase.
On decompression, the HP-I phase remains down to 2 GPa, displaying a large hysteresis in $V$$_{f.u.}$ and $c$ below 12 GPa.
Considering the significant hysteresis observed in $R$$_{xx}$ vs. $P$, the $C2/m$ $\rightarrow$ HP-I transition is considered first-order.
We later discuss those anomalous compressions concerning electronic transformations, the HS to LS crossover, and the volume collapse.

\begin{figure}
\includegraphics[scale=0.26]{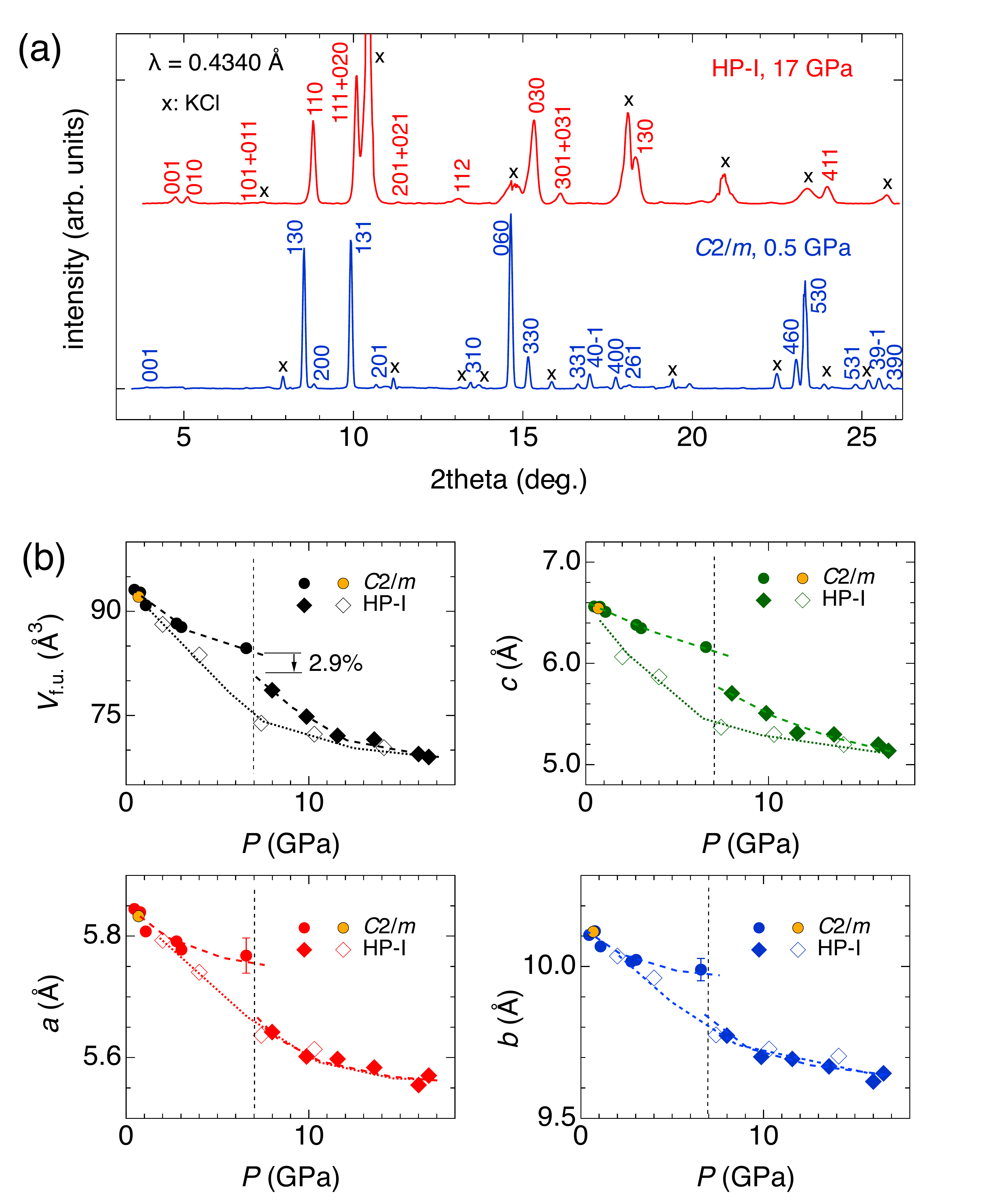}%
\caption{\label{fig2}(color online) Structure analysis of CoPS$_3$ by XRD under pressure. (a) XRD patterns of the $C2/m$ (0.5 GPa) and HP-I (17 GPa) phases. Note that KCl changes from B1 to B2 type structure at 2 GPa. (b) Pressure evolution of \textit{V}$_{f.u.}$ and lattice constants. The trigonal structure of HP-I is reduced to a monoclinic lattice ($\beta$ = 90$^{\circ}$) for comparison. The open and closed data points indicate compression and decompression, respectively. The error bars are inside the data markers at most pressures except for 6.8 GPa. The broken and dashed curves are guides for the eyes. The vertical dotted line is the structural phase boundary.}
\end{figure}

\subsection{Raman scattering}
To obtain further insights into changes to the symmetry of the crystal structure under compression, we performed pressure-dependent Raman scattering measurements.
Fig. 4a displays the evolution of the Raman spectra from CoPS$_3$ under quasi-hydrostatic compression at room temperature.
Based on the previous theoretical and experimental reports, we anticipate eight Raman active modes (5E$_g$ + 3A$_g$) for bulk CoPS$_3$ at room temperature \cite{Liu2021}.
At pressures below 6.7 GPa (in the stability region of $C2/m$), we observe all eight peaks in agreement with the previous reports \cite{Liu2021}.
The peak at 110 cm$^{-1}$ is not observed at atmospheric pressure but becomes visible above 1.2 GPa, suggesting that a preferential alignment of layers could cause this peak to be more prominent at higher pressures.
Up to 6.7 GPa, all the peaks blueshift in frequency with increasing pressure, as expected for phonon modes under compression.
Between 6.7 GPa and 7.5 GPa, the Raman spectrum changes abruptly with the loss of peaks and the appearance of new peaks (Fig. 3a and 3b).
Accompanied by the Raman spectral change, the sample becomes lighter in color and more reflective under epi-illumination (Fig. 3a, photographs), in agreement with the visible observations during our transport measurements (Fig. 1a). 
The abrupt Raman spectra and reflectivity changes give further evidence of the $C2/m$ $\rightarrow$ HP-I structural change coinciding with the ITM transition. 
Similar to what our electrical resistance measurements observe, the transition in the Raman spectra takes place within a narrow pressure range of ~0.8 GPa, strongly suggesting the absence of an intermediate phase between $C2/m$ and HP-I phases.
At 18.7 GPa, all peaks diminish significantly and are replaced by a broad peak between 300-400 cm$^{-1}$, except for the peak near 120 cm$^{-1}$.
We consider two possibilities for the cause of broadening.
One is the solidification of the pressure-transmitting medium near 10 GPa, and the developing non-hydrostatic condition that induces the inhomogeneous strain in the crystal \cite{Klotz2009, Fujishiro1982, Tateiwa2010}.
Another possibility could be the indication of further structural transformation.
Future studies would address the question.

Figure 3b, left panel, shows the pressure-dependent frequencies of the Raman peaks.
The right panel in Fig. 3b shows a magnified view of the ambient pressure phase. 
Based on the previous literature, we assign the peak P$_4$ near 245 cm$^{-1}$ to the out-of-plane bending (A$_{1g}$) of P$_2$S$_6$$^{4-}$ dimer units. 
It has the highest pressure coefficient, reflecting the large compressibility in the $c$-axis (Fig. 3b) \cite{Liu2021}. 
P$_8$ near 560 cm$^{-1}$ is the out-of-plane stretching of P-P dimer (E$_g$) \cite{Liu2021}.
The peak shows a sharp increase in the frequency at the start of compression.
Contrary, P$_6$ near 380 cm$^{-1}$ representing in-plane stretching of P$_2$S$_6$$^{4-}$ units (A$_{1g}$) has the lowest pressure coefficient \cite{Liu2021}.
Overall, the compression affects out-of-plane phonon modes more than in-plane modes.

\begin{figure}
 \includegraphics[scale=0.46]{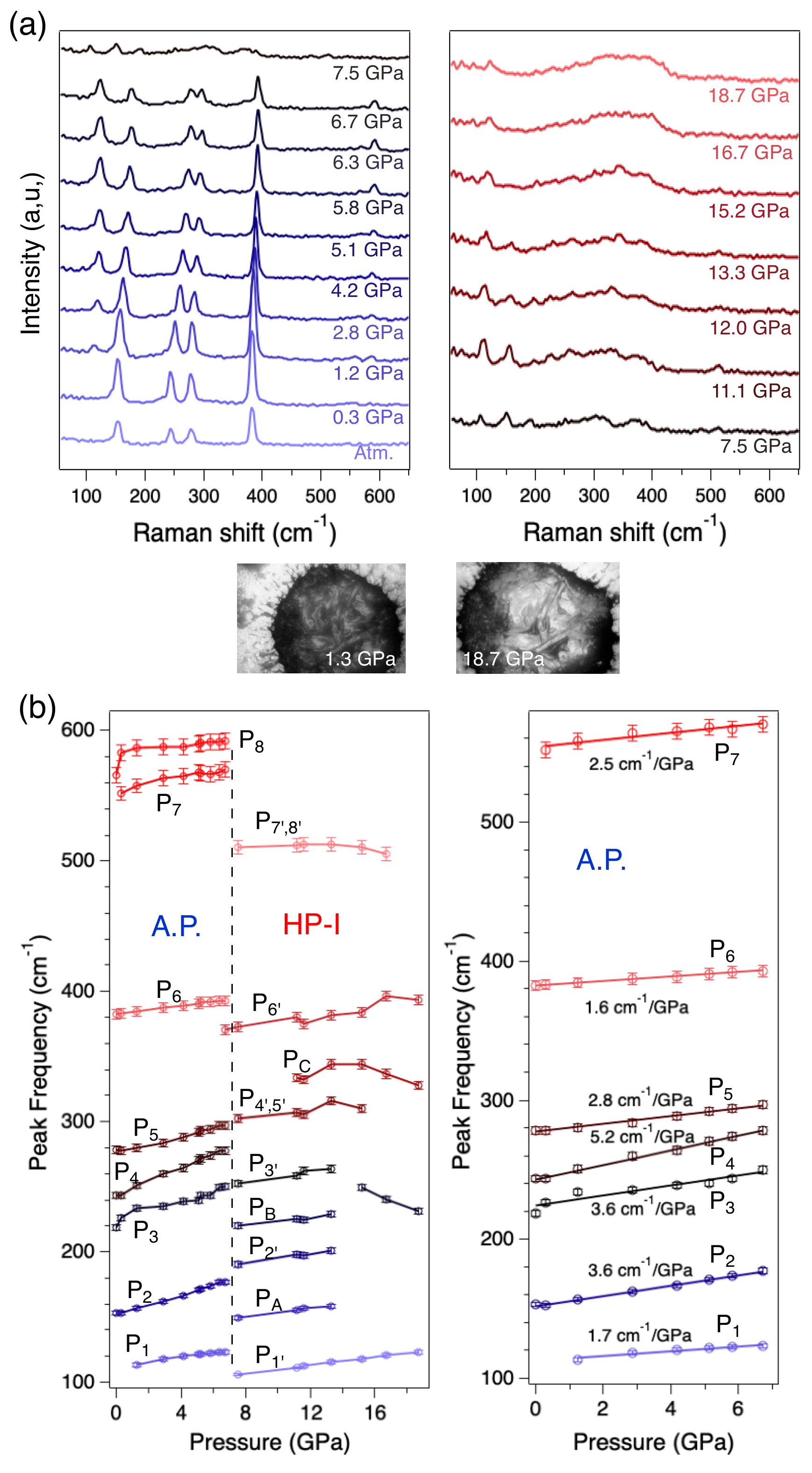}%
 \caption{\label{fig3}(color online) Raman scattering measurements of CoPS$_3$ under pressures. (a) The pressure evolution of the Raman scattering spectra at ambient temperature. Pictures show CoPS$_3$ at 1.3 GPa and 18.7 GPa. Atm.; atmospheric pressure (b-left) Pressure dependencies of the Raman scattering peaks. Dashed vertical lines show the structural phase boundaries. (b-right) The close-up view of the left panel. See the Supplemental Material for the peak-fitting analysis \cite{supp}.}
\end{figure}

Next, we try to understand the $C2/m$ $\rightarrow$ HP-I transition.
We first track the peak frequency changes across the structural transition.
P$_1$ near 115 cm$^{-1}$ (complex vibrations along all three axes of Co and S, E$_g$) blue shifts with pressure up to 7 GPa, above which it redshifts in frequency (labeled P$_{1'}$ in Fig. 3b).
P$_2$ near 145 cm$^{-1}$ (in-plane Co-Co stretching, E$_g$) and P$_3$ near 238 cm$^{-1}$ (out-of-plane stretching of P$_2$S$_6$$^{4-}$, E$_g$) behave similarly to P$_1$, blueshifting up to 7 GPa, followed by slight discontinuities and are designated as P$_{2'}$ and P$_{3'}$in the HP-I phase, respectively. 
P$_4$ near 240 cm$^{-1}$merges with P$_5$ near 280 cm$^{-1}$ (in-plane S-S vibration in P$_2$S$_6$$^{4-}$ units, E$_g$) up to 7 GPa. 
In the HP-I phase, the merged peak is designated P$_{4',5'}$ near 300 cm$^{-1}$.
P$_6$ near 380 cm$^{-1}$ (in-plane S-S vibration, A$_{1g}$) exhibits a sharp discontinuity across the phase transition and appears at a lower frequency ~400 cm$^{-1}$ labeled as P$_{6'}$.
P$_7$ near 545 cm$^{-1}$ (complex stretching mode of P$_2$S$_6$$^{4-}$, E$_g$) and P$_8$ come closer up to 7 GPa.
Across the critical pressure, these peaks disappear, and the highest frequency peak in HP-I appears at a much lower frequency ~510 cm$^{-1}$, designated as P$_{7',8'}$.
The merging and newly appeared peaks provide evidence for both increasing and decreasing symmetry.

We next consider the evidence for increased symmetry.
As mentioned above, several Raman peaks merge with increasing pressure up to 7 GPa (P$_4$+P$_5$ $\rightarrow$ P$_{4',5'}$, and P$_7$+P$_8$ $\rightarrow$ P$_{7',8'}$).
This means that in addition to the mirror planes, axial glides, inversion centers, two-fold rotations, and the two-fold screw axes that characterize the $C2/m$ space group, the system gains additional symmetry elements across 7 GPa. 
Examination of the $C2/m$ group $\leftrightarrow$ supergroup relationships provides several higher symmetry candidates such as $P\overline{3}1m$, $P\overline{3}m1$, and $Cmmm$. 
Since our XRD result suggests a trigonal structure, we can exclude $Cmmm$.
See the Supplemental Material of Ref. \cite{Harms2022} for the summarized subgroup/supergroup symmetry relations relevant to the $M$PS$_3$ compounds.

Next, we investigate the evidence for symmetry breaking.
The newly emergent peak P$_A$ (near 140 cm$^{-1}$) in the HP-I phase indicates the rise of the out-of-phase intralayer Co translational mode.
P$_B$ and P$_C$ are considered to be related to the S-S vibrations from their frequencies in the lower-pressure phase.
To unravel how these modes correspond to the change in crystal symmetry, we consider the relevant mode displacement patterns and how they impact different symmetry elements. 
These include some newly established symmetry elements of $P\overline{3}1m$ and $P\overline{3}m1$, such as axial glide planes, two-fold rotations, and two-fold screw axes. 
Therefore, candidate subgroups include $P31m$, $P\overline{3}$, $P3m1$, and $P312$. 
In the Raman spectra, we do not see a significant increase in the overall number of peaks, suggesting the retention of the inversion center. 
Of the four candidate subgroups, $P\overline{3}$ retains the inversion center. 
We, therefore, conclude the $C2/m$ $\rightarrow$ $P\overline{3}$ transformation. 

Figure 4 summarizes the crystal structures of CoPS$_3$ in the $C2/m$ and $P\overline{3}$.
Note that we did not refine the atomic positions in the $P\overline{3}$ due to the limitation mentioned above in the obtained XRD data \cite{supp}.

We calculated the mode Gr\"{u}neisen parameter for each phonon mode using the pressure dependence of the Raman frequencies and the $V$$_{f.u.}$.
See the Supplemental Materials for the analysis and results \cite{supp}.

 \begin{figure}
 \includegraphics[scale=0.44]{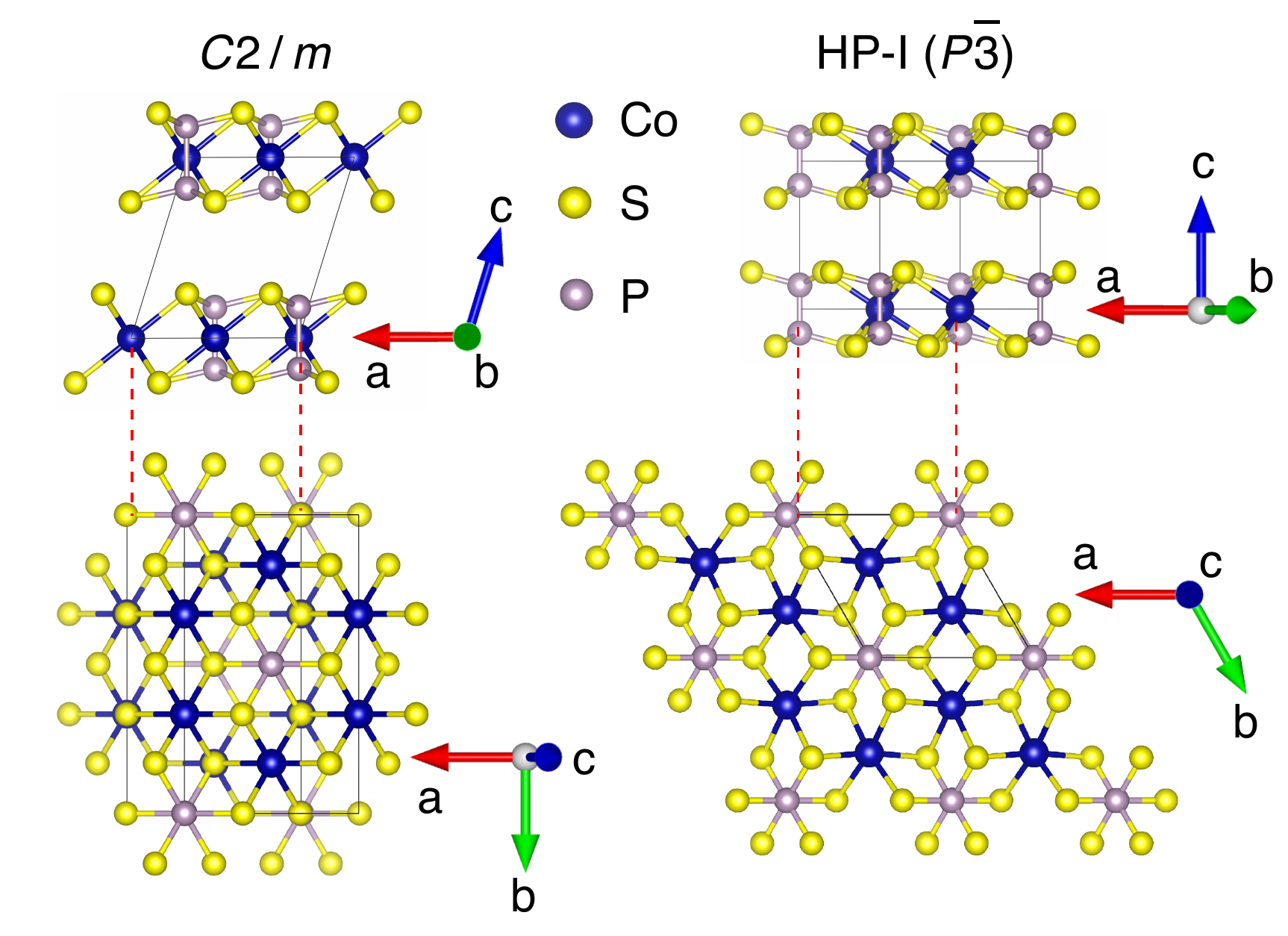}%
 \caption{\label{fig4}(color online) The crystal structure of A.P.-phase ($C2/m$) and the suggested $P\overline{3}$ structure for HP-I phase visualized using VESTA\cite{Momma2011}.}
\end{figure}

\subsection{Transverse transport measurements}
Now that we have confirmed the existence of a new metallic phase at high pressure in CoPS$_3$, we turn our attention to its electronic properties.
Transverse transport measurements, including Hall resistance and magnetoresistance (MR), are essential probes because it gives us information about the Fermi surface \cite{Savary2018}.
Fig. 5a and Fig. 5b display the Hall resistivity ($\rho$$_{xy}$) and MR ($\Delta$$\rho$$_{xx}$($B$)$/$$\rho$$_{xx}$(0)) of sample \#1 at 10, 13, and 15 GPa and temperatures ranging from 1.8 to 160 K.
The $\rho$$_{xy}$ shows positive and mostly linear magnetic field dependence at all temperatures and pressures (Fig. 5a). 
Besides, it does not saturate at this study's highest field (9 T).
If a material is ferromagnetic, the anomalous Hall effect is anticipated with saturation at a high field.
Our Hall resistance data thus indicate that CoPS$_3$ is not ferromagnetic. 
CoPS$_3$ is likely a hole-dominant conductor that requires a multiple-conduction band model. 
We also notice that $\Delta$$\rho$$_{xx}$($B$)$/$$\rho$$_{xx}$(0) exhibits a positive and linear magnetic field dependence (linear magnetoresistance: LMR) at all pressures and temperatures (Fig. 5b).

Generally, a multi-band analysis based on a semi-classical description utilizes information such as carrier density and mobility on the Fermi surface from the Hall resistivity and MR.
However, the obtained results tend to become ambiguous because of hypothesizing the number of carrier types.
The LMR in this study makes the analysis even more complicated because it is far from the quadratic behavior anticipated by a semi-classical description.
Here, we employ a simple one-band model analysis and estimate the orders of density ($n$$_{e,av.}$) and mobility ($\mu$$_{e,av.}$), on average, for all carriers.
From the relation $1/eR$$_H$ = $n$$_{e,av.}$, the Hall coefficient (\textit{R}$_{H}$) and the electron charge ($e$) provide the estimation $n$$_{e,av.}$ = 1.41$\times$10$^{21}$ cm$^{-3}$ at 10 GPa and 1.8 K.
This value is slightly smaller than that a Hall effect measurement expects for general metals (10$^{22}$ cm$^{-3}$).
Using the relation $\sigma$ = $\mu$$_{e,av.}$$n$$_{e,av.}$, the $\mu$$_{e,av.}$ = 138 cm$^2$$V$$^{-1}$$s$$^{-1}$ can be extracted, where $\sigma$ is a conductivity.
The $n$$_{e,av}$ and $\mu$$_{e,av}$ are within the range of 1.4 - 2.9$\times$10$^{21}$ cm$^{-3}$ and 55 - 138 cm$^2$$V$$^{-1}$$s$$^{-1}$ at 2 K for all pressures, respectively.

Here, we analyze the observed LMR (Figs. 5b and 5c).
The semi-classical model predicts that the $\rho$$_{xx}$ evolves quadratically with a magnetic field, saturating at high fields if the hole and electron densities are not compensated \cite{Lifshitz1957, Ziman1958}.
At the low-field limit, $\omega$$_{c}$$\tau$$\ll$ 2$\pi$, where $\omega$$_{c}$  is the cyclotron frequency and $\tau$ is the relaxation time, the leading term in  $\rho$$_{xx}$ becomes $\Delta$$\rho$ = $\rho$($B$)$-$$\rho$($B$=0) $\sim$ $H$$^2$  due to Onsager reciprocity relation, which requires $\sigma$$_{ij}$(\textit{\textbf{B}}) = $\sigma$$_{ji}$($-\textit{\textbf{B}}$) \cite{Lifshitz1957, Ziman1958}.
The $\rho$$_{xx}$ of CoPS$_3$ evolves almost linearly with the field contrary to the semi-classical description.
The LMR is observable down to low fields: 0.7 T at 10 GPa, 2 T at 13 GPa, and 3 T at 15 GPa (Fig. 5c) followed by the asymptotic curves approaching zero near zero field.
Elevated temperature suppresses the increase of the $\rho$$_{xx}$ vs. $B$ curves.
At higher pressures, the parabolic shape in the $\rho$$_{xx}$ vs. $B$ becomes more evident at lower fields.
To obtain further insight into the linear term in the $\rho$$_{xx}$ vs. $B$ relation, we adopt a phenomenological approach to disentangle these components, fitting the measured MR as $\rho$$_{xx}$($H$, various $T$) = $\rho$$_{xx}$($T$, $H = 0$) + $A$($T$)$H$ + $B$($T$)$H$$^2$ \cite{Feng2019}.
We perform the fitting below 4 T where the quadratic component is visible.
Fig. 5d plots $A(T)$ and $B(T)$ as a function of temperature.
The $A(T)$ saturates below 10 K, significantly decreasing with increasing temperature.
At low temperatures, the $A(T)$ decreases with pressure from 10 to 13 GPa.
However, the change becomes diminished between 13 and 15 GPa. 
By comparison, the $B(T)$ does not change appreciably over a wide range of temperatures.
We try to discern the origin of LMR in the following section.
\begin{figure*}
 \includegraphics[scale=0.38]{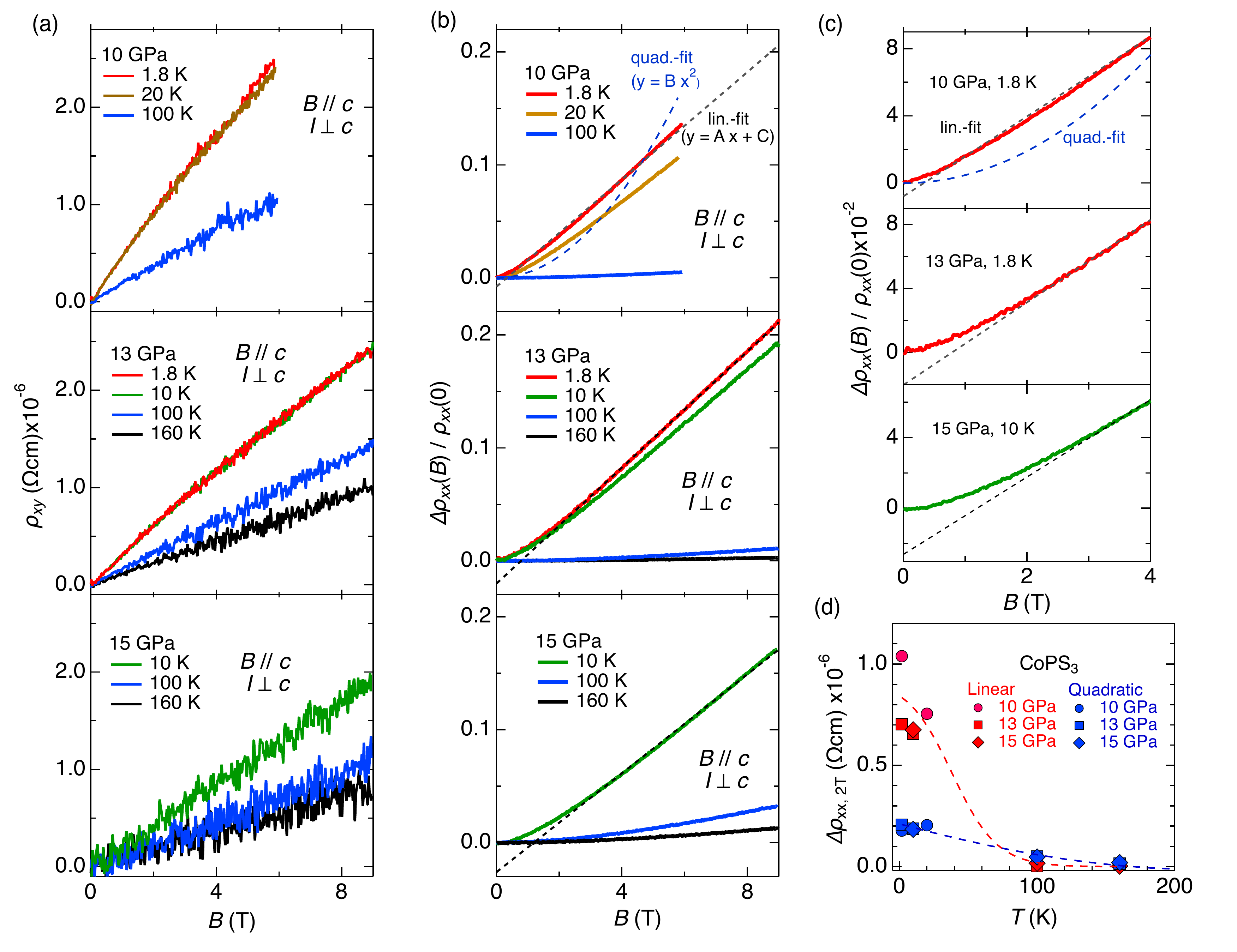}%
 \caption{\label{fig5}(color online)  Transverse transport data from metallic CoPS$_3$ at 10, 13, and 15 GPa. (a) The $\rho$$_{xy}$ at different temperatures. The data at 1.8 K and 20 K at 10 GPa overlap (top panel). The data at 1.8 K and 10 K at 13 GPa overlap (middle panel).  (b) The $\Delta$$\rho$$_{xx}$($B$)/$\rho$$_{xx}$($B=0$). The dotted (black) and dashed (blue) lines in the top panel indicate the linear fit, ${y = A x + C}$ with $A =$ 1.927(3)$\times$10$^{-3}$, $C = -$7.7772(3)$\times$10$^{-3}$) and quadratic-fits (quad.-fit, ${y = Bx}$$^2$ with ${B = }$4.82(4)$\times$10$^{-3}$) to the data at 10 GPa and 1.8 K.  (c) $\Delta$$\rho$$_{xx}$($B$)/$\rho$$_{xx}$($B=0$) at low field. (d) The linear and quadratic components of MR under a field of 4 T for CoPS$_3$. The broken lines are a guide for eyes.}
\end{figure*}

\begin{table*}[htb]
\caption{Structural and electronic phase diagram of $M$PS$_3$s that metalize under compression. AFM$_{out}$ and AFM$_{in}$ mean that the magnetic moments are pointing mostly out-of-plane and in-plane directions, respectively. The \textit{P}$_c$ is the pressure where crystal structures start transformations. $\Delta$$V$ is the volume reduction at  insulator-to-metal transition.}
\begin{ruledtabular}
 \begin{tabular}{lllllll}
        & A.P. & H.P.  &  &   &  & Ref. / Note \\
       \hline\\
    V$_{0.9}$PS$_3$  & $C2/m$  ($\beta$$\sim$107$^{\circ}$)  & $C2/m$ ($\beta$$\sim$90$^{\circ}$) & & &  & \cite{Coak2019, Ouvrard1985V}\\
    & AFM$_{NA}$\footnotemark[1], $\textit{\textbf{q}}=[010]$, ins. &ins.\footnotemark[2] & met.\footnotemark[3] & & &\\
    \hspace{6 mm}\textit{P}$_c$& $-$ & 2.6 GPa &12 GPa &&&\\
        &&&$\Delta$$V$$\sim$0.8\%&&&\\
   \hline\\
   
    MnPS$_3$ & $C2/m$ ($\beta$$\sim$107$^{\circ}$) & $P\overline{3}1m$ & $C2/m$ ($\beta$$\sim$90$^{\circ}$)  & &  & \cite{Ourvrard1985, Harms2020, Wang2016a,Coak2020}\\
    & AFM$_{out}$, $\textit{\textbf{q}}=[010] or [000]$, ins.& ins. & met. &   & &spin crossover\\
   \hspace{6 mm}\textit{P}$_c$ &$-$&10 GPa&28 GPa, $\Delta$$V$$\sim$19.7\% &&&at 28 GPa\\
    &&&&&&\\
       \hline\\

    FePS$_3$& $C2/m$ ($\beta$$\sim$107$^{\circ}$) & $C2/m$ ($\beta$$\sim$90$^{\circ}$) & $P\overline{3}1m$  &  &  & \cite{Ourvrard1985, Wang2018a,Coak2021,Coak2020, Harms2022a} \\
    & AFM$_{out}$, $\textit{\textbf{q}}=[01\frac{1}{2}]$, ins. & AFM, $\textit{\textbf{q}}=[010]$, ins. & met. & & &spin crossover\\
    \hspace{6 mm}\textit{P}$_c$ &$-$&2 GPa&14 GPa, $\Delta$$V$$\sim$10.6\%&&&at 14 GPa\\
    &&&&&&\\
           \hline\\

    CoPS$_3$ & $C2/m$ ($\beta$$\sim$107$^{\circ}$)  & $P\overline{3}$ &  &&  &\cite{Ourvrard1985, Wildes2017}\\
    & AFM$_{in}$, $\textit{\textbf{q}}=[010]$, ins.&  met. & & &&this study\\
    \hspace{6 mm}\textit{P}$_c$ &$-$ &7 GPa, $\Delta$$V$$\sim$2.9\%&&&&\\
      &&&&&&\\
        \hline\\
   
     NiPS$_3$ & $C2/m$ ($\beta$$\sim$107$^{\circ}$) &   $P\overline{3}$ & $P\overline{3}1m$ &  $P3m1$ &&\cite{Ourvrard1985, Harms2022, Ma2021, Matsuoka2021} \\
    & AFM$_{in}$, $\textit{\textbf{q}}=[010]$, ins. & ins. &ins. & met.& &\\
    \hspace{6 mm}\textit{P}$_c$&$-$&7.2 GPa& 15 GPa& 23 GPa&\\
      &&&&$\Delta$$V$$\sim$2.8\%&&\\
   \hline\\

    CdPS$_3$  & $C2/m$ ($\beta$$\sim$107$^{\circ}$)  & $R\overline{3}$ & $R\overline{3}$ &&  &\cite{Ourvrard1985, Niu2022} \\
    &ins.&&&&&\\
    \hspace{6 mm}\textit{P}$_c$&$-$ & 0.25 GPa& 8.7 GPa& & &\\
  \end{tabular}
  \end{ruledtabular}
            \footnotetext[1]{N.A.: information Not-Available} \footnotetext[2]{ins.: insulator} \footnotetext[3]{met.: metal}
\label{tb:lattice}
\end{table*}

\section{Discussion}
Here, we discuss the electronic configuration of the metallic CoPS$_3$, discerning the source of the observed pressure dependence of $V$$_{f.u.}$ and the LMR.
First, we focus on the $V$$_{f.u.}$ reduction at $C2/m$ $\rightarrow$ $P\overline{3}$ transformation.
As discussed earlier, we concluded HS $\rightarrow$ LS spin crossover takes place at the insulator-to-metal transition and structural transformation.
However, the relatively small reduction in the $V$$_{f.u.}$ is not explained simply by the decrease in the ionic radius of Co$^{2+}$.
Remarkably, the theoretical study predicts that the magnetic moment in CoPS$_3$ is much more robust than Fe$^{2+}$ and Mn$^{2+}$ under pressure \cite{Gu2021}.
The study suggests that the CoPS$_3$ in either the $R\overline{3}$ or $C2/m$ ($\beta$$\sim$90$^{\circ}$) phases above 12.5 GPa is ferromagnetic. The magnetic moments of Co$^{2+}$ decrease significantly with increasing pressure but do not achieve $S=1/2$ even at 50 GPa \cite{Gu2021}. 
Although the predicted crystal structure differs from the one our experiments determine, we then raise the possibility that the $P\overline{3}$ phase is in the middle of spin crossover where the HS- and LS-Co$^{2+}$ coexist.

We turn our eyes to the pressure dependence of the $V$$_{f.u.}$ in the $P\overline{3}$ phase.
Looking at several Fe-bearing compounds and (Mg, Fe)O forsterite, we find the changes in the pressure dependence of the volume due to the interplay between the compressibility and spin variation effect on the Fe$^{3+}$ ionic radius \cite{Chen2012, Greenberg2017, Rozenberg2005}.
Considering the incomplete spin crossover discussed above, the significant reduction of $V$$_{f.u.}$ between 7 and 12 GPa is potentially due to the proceeding HS $\rightarrow$ LS crossover.
Then, the moderate $V$$_{f.u.}$ vs. $P$ slope above 12 GPa suggests the spin crossover's completion or moderate progress.
The significant hysteresis between the $V$$_{f.u.}$ of $P\overline{3}$ upon compression and decompression, especially below 12 GPa, could be because the sample on compression has a bigger fraction of HS than decompression.

Next, we try to discern the source responsible for LMR.
LMR has been observed in a growing number of novel materials and often invoked as evidence for some exotic quasiparticles in materials \cite{Wang2011b, Zhao2013, Tang2011, Wang2012e, Gusev2013, Wang2012b, Novak2015, Liang2015, Zhao2015a, Feng2015a, Shekhar2015, Zhao2015, Xu1997, Zhang2011, Sinchenko2017, Feng2019, Khouri2016, Gusev2013}.
At the high-field limit $\omega$$_{c}$$\tau$ $\gg$ 2$\pi$, where $\omega$$_{c}$ is the cyclotron frequency, and the $\tau$ is the relaxation time, there have been several suggested electronic and geometric mechanisms that satisfy the criteria for a quantum LMR.
The first is special features on the Fermi surface, including the linear dispersion from a Dirac cone with infinitesimally small carrier mass \cite{Abrikosov1998, Abrikosov2000, Young1968, Naito1982}.
The second is principally geometric in nature, including an average over a combination of open and closed electron trajectories in polycrystals \cite{Ziman1958, Song2015, Parish2003, Xu1997, Hu2008}.

On the other hand, the disorders of density and spin have been suggested as universal mechanisms.
The density disorder provides an inhomogeneous distribution of charge concentration and affects the conduction carrier path, admixing the Hall resistance component with MR \cite{Abrikosov2000, Khouri2016, Hu2008}.
Similarly, the LMR due to the magnetic disorder has been observed for several 3$d$ ferromagnets and the antiferromagnetic normal conducting state of FeSe \cite{Raquet2001, Wang2017}.
Another is the LMR in CDW and SDW containing materials \cite{Feng2019}.
From their nature, those three are applicable to the LMR to low-filed limit $\omega$$_{c}$$\tau$ $\ll$ 2$\pi$.

To examine the suggested mechanisms, we first estimate an average for all carriers $\omega$$_{c}$$\tau$ = $B$/$\rho$$nec$ = 2.75$\times$10$^{-2}$ at 2 T (10 GPa, 1.8 K) for CoPS$_3$.
The criteria for quantum LMR ($\omega$$_{c}$$\tau$ $\gg$ 2$\pi$) (Ref. \cite{Abrikosov2000}) are thus not satisfied under our measurement conditions.
Additionally, CDW and SDW are not likely, judging from the featureless $\rho$$_{xx}$ vs. $T$.
Also, we do not observe the appearance of satellite peaks in XRD, which is suggestive of CDW.
Finally, the rapid diminution of $A(T)$ (Fig. 5d) at elevated temperatures argues against the phonon- \cite{Young1968} or the excitation-based \cite{Sinchenko2017} scattering mechanisms.

We next test the density fluctuation scenario.
Since our sample is under non-hydrostatic stress, it is the most straightforward one to consider.
However, the estimated $n$$_{e,av.}$ and $\mu$$_{e,av.}$ are far bigger and smaller than that of the high-mobility and low-carrier density materials where the density fluctuation effects become more prominent \cite{Khouri2016, Pan2005, Simon1994, Hu2008}.
Besides, since the pressure gradient in the sample generally develops with pressure in a non-hydrostatic condition, a more amplified density fluctuation and even more linear MR are expected, contrary to our experimental results.
Thus, we defer concluding the density fluctuations as the dominant source.

Finally, we consider the spin-disordered mechanism \cite{Raquet2001, Wang2017}.
In this model, ions with different magnetic moments coexist randomly.
The inhomogeneously distributed magnetic moments possibly affect the conduction carrier trajectories allowing irregular current paths and the LMR.
Based on our conclusion of the incomplete HS $\rightarrow$ LS spin crossover, it can be thought that HS- and LS-Co$^{2+}$ ions coexist and are arranged in a disordered manner, possibly possessing a short-range magnetic ordering.
In Fig. 5c and 5d, we see the quadratic component of the LMR becomes more evident in the $\rho$$_{xx}$ vs. $B$ at higher pressure, implying a more homogeneous magnetic moment distribution promoted by pressure.
Besides, the linear component at 13 GPa and 15 GPa possess almost the same value.
Those observations are consistent with the interpretation that the spin crossover proceeds with pressure up to 12 GPa and stops or progresses moderately above 12 GPa in the $P\overline{3}$ phase.

The remaining question is how the insulator-to-metal transition, the structural transition, and the spin crossover relate to each other in CoPS$_3$.
We propose two scenarios.
The first is that the insulator-to-metal transition occurs simultaneously with the structural transition.
The second is that metallization originates within the $P\overline{3}$ phase.
This scenario arises from the observation that CoPS$_3$ is still a semiconductor at 7.4 GPa, while our Raman and XRD measurements suggest that the structural transformation occurs at 7 GPa and completes within 1 GPa.
Besides, upon decompression, CoPS$_3$ reverts to an insulator at 3.5 GPa, preceding the $P\overline{3}$ $\rightarrow$ $C2/m$ transition observed below 2 GPa.
Currently, we do not have precise and detailed data to address the discrepancy in the transition pressures.
The off-stoichiometry of the sample may also affect the transition pressures.
Future detailed studies including the simultaneous measurements of electrical transport and crystal structure would provide an unambiguous answer.

Table 1 summarizes the structural and electronic evolution of $M$PS$_3$ ($M$ = V, Mn, Fe, Co, Ni, Cd) reported to date.
All $M$PS$_3$ exhibit $C2/m$ ($\beta$$\sim$107$^{\circ}$) to trigonal structural transition when subjected to pressure, decreasing their monoclinic angle to 90$^{\circ}$ as a consequence of the inter-layer sliding.
The insulator-to-metal transition commonly occurs when $M$PS$_3$ compounds are in the trigonal or $C2/m$ with $\beta$$\sim$90$^{\circ}$ symmetries. 

The question is how the electronic configuration, magnetism, and structure correlate.
We see that $M$PS$_3$ compounds, as far as the available experimental data display, can be classified into two groups concerning their pressure transformations.
The first contains MnPS$_3$ and FePS$_3$ and exhibits the transformation between symmetries in the group-subgroup relation ($C2/m$ $\leftrightarrow$ $P\overline{3}1m$).
These compounds align their magnetic moments mainly in the out-of-plane direction.
The second group is comprised of CoPS$_3$ and NiPS$_3$.
These materials have moments aligned largely in-plane.
Contrary to the first group, the transition process from the lower-pressure phase ($C2/m$) to the first high-pressure phase is not in the simple group-subgroup relation.
As discussed, CoPS$_3$ transforms from $C2/m$ (14) to $P\overline{3}$ (147) via a higher-symmetry phase such as $P\overline{3}1m$ (162). 
NiPS$_3$ is an insulator in the $P\overline{3}$ phase.
Thus, Ni$^{2+}$ ions in the $P\overline{3}$ phase may possibly be in the HS state.
The potentially remaining magnetic moments of Co$^{2+}$, and possibly Ni$^{2+}$, are likely to affect the high-pressure phase that succeeds the $C2/m$.

Figure 6 is the visual summary of the $T$$_N$s and $V$$_{f.u.}$ shown in Table I as a function of the ionic radius of $M$ in HS states.
In Fig. 6, the two groups (Mn and Fe, Co and Ni) discussed for the pressure-induced structural transition are noticeable in the $T$$_N$ and $V$$_{f.u}$.
The $V$$_{f.u.}$s of NiPS$_3$ and CoPS$_3$ at 1 bar are similar.
However, the $V$$_{f.u.}$ shows an obvious increase from FePS$_3$ to MnPS$_3$ while the ionic radius difference between Fe$^{2+}$ and Mn$^{2+}$ is smaller than between Ni$^{2+}$ and Co$^{2+}$.
The $T$$_N$ decreases moderately with ionic radius from Ni$^{2+}$ to Co$^{2+}$, followed by a steep decline from Fe$^{2+}$ to Mn$^{2+}$.
Remarkably those two groups are also observable in the volume reduction from the insulator to the metallic phases at the metallization pressure.
NiPS$_3$ and CoPS$_3$ commonly show a smaller reduction than another group (Fe and Mn).
In $M$PS$_3$ compounds, the $t$$_{2g}$ and $e$$_g$ orbitals play roles in bonding, and the hopping integrals between $t$$_{2g}$ and $e$$_g$ show different anisotropy \cite{Kim2019b}.
Thus, the occupation of the electron orbitals, partially filled $t$$_2g$ and $e$$_g$ orbitals of FePS$_3$ and filled $t$$_{2g}$ in NiPS$_3$ for example, has a direct influence on the physical characteristics, making this group of compounds a rich platform to explore novel quantum phenomena.
Fig. 6 reveals that the electronic configurations at ambient pressure influence the high-pressure properties.
We also notice that MnPSe$_3$ and FePSe$_3$ possess similar $T$$_N$ with their analogous MnPS$_3$ and FePS$_3$.
If Ni and Co follow this trend, we may expect $T$$_N$s of NiPSe$_3$ and CoPSe$_3$ to be close to NiPS$_3$ and CoPS$_3$.

\begin{figure}
 \includegraphics[scale=0.42]{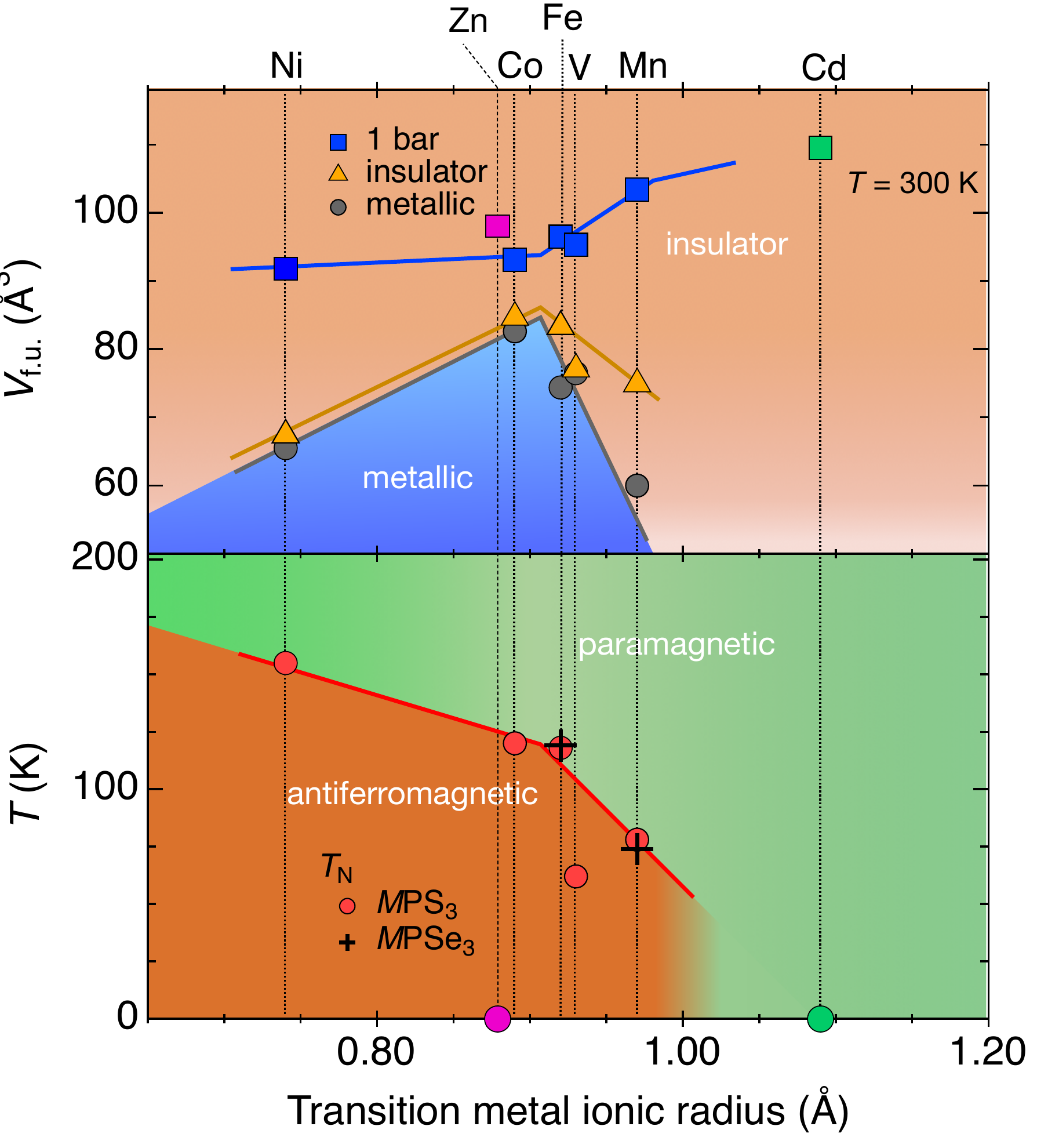}%
 \caption{\label{fig6}(color online) $M$PS$_3$ phase diagram with temperature and volumes along the vertical axes and transition metal ($M$$^{2+}$) ionic radius in HS state (except for V) along the horizontal axis. Data are taken from Ref.\cite{Coak2019, Ouvrard1985V, Ourvrard1985, Harms2020, Wang2016a,Coak2020, Wang2018a,Coak2021, Wildes2017, Ma2021, Matsuoka2021, Niu2022, Wang2018a, Shannon1976, Prouzet1986}. The solid marks in the upper panel indicate the $V$$_{f.u.}$ at 1 bar (square), insulator phases (triangle), and metallic phases (circle). In the lower panel, the $T$$_N$ of $M$PS$_3$ are plotted with MnPSe$_3$ and FePSe$_3$.}
\end{figure}

\section{\label{sec:level4}SUMMARY}
We have successfully grown high-quality single crystals of CoSP$_3$ suitably large enough for conducting various high-pressure experiments.
We studied the electrical transport and structural evolution of CoPS$_3$ under quasi-uniaxial pressure along the layer-stacking direction through electrical resistance, Hall resistance, magnetoresistance, Raman scattering, and XRD measurements.
Electrical resistance significantly decreases as the pressure increases, consistent with the rise of the optical reflectivity of the sample.
CoPS$_3$ becomes metallic above 7 GPa, accompanied by the monoclinic $C2/m$ $\rightarrow$ trigonal $P\overline{3}$ structural transition.
Metallic CoPS$_3$ shows no superconducting transition down to 2 K.
The $C2/m$ $\rightarrow$ $P\overline{3}$ transformation induces a 2.9\% reduction in $V$$_{f.u.}$, much smaller than that of the Mn and Fe analogous.
The Hall effect data indicate the metallic CoPS$_3$ is a hole-dominant conductor.
We observed the linear magnetoresistance in a wide range of magnetic fields.
The linear magnetoresistance, the small volume reduction across the structural transition, and the previous theoretical prediction \cite{Gu2021} suggest the coexistence of HS- and LS-Co$^{2+}$ ions and the inhomogeneous magnetic moment distribution with a possible short-range magnetic ordering.
Furthermore, the anomalous compression behavior of $V$$_{f.u.}$, and the pressure evolution of the electrical resistance and linear magnetoresistance suggest the possibility that the metallization occurs within the $P\overline{3}$ and the spin crossover completes up to 12 GPa, or the progresses becomes moderate above the pressure in the $P\overline{3}$.
By revealing the high-pressure phase and electrical transport property of CoPS$_3$, this report summarizes the comprehensive phase diagram of $M$PS$_3$ ($M$ = V, Mn, Fe, Co, Ni, Cd) that metalize under compression.
$M$PS$_3$ at ambient pressure has been an excellent platform for exploring emergent quantum phenomena due to their various electronic configurations.
The phase diagram reveals that the electronic configurations at ambient pressure strongly influence the structural and electronic properties of $M$PS$_3$ at high pressures, making this series of compounds suitable platforms for exploring new physical properties under compression.

\begin{acknowledgments}
This research is funded by the Gordon and Betty Moore Foundation’s EPiQS Initiative, Grant GBMF9069 to D.M.
XRD experiments in this study were performed at GSECARS (Sector 13), Advanced Photon Source (APS), Argonne National Laboratory. 
GSECARS is supported by the National Science Foundation – Earth Sciences (EAR-1634415).
This research used resources of the Advanced Photon Source, a U.S. Department of Energy (DOE) Office of Science User Facility operated for the DOE Office of Science by Argonne National Laboratory under Contract No. DE-AC02-06CH11357. Crystal growth and Raman characterization were performed under Air Force Office of Scientific Research (AFOSR) grant LRIR 23RXCOR003 and AOARD-MOST Grant Number F4GGA21207H002.
Part of the XRD experiment was supported by COMPRES under NSF Cooperative Agreement EAR-1606856.
We thank Dr. Antonio M. dos Santos and Prof. Maik Lang for providing us with the DACs for the XRD measurements.
We thank Prof. Heung-Sik Kim and Dr. Sungmo Kang for the fruitful comments and discussions.
\end{acknowledgments}

%

\end{document}